\documentclass[printer]{aa}  
\usepackage[dvipsnames, svgnames]{xcolor}
\usepackage[colorlinks=true, citecolor=Navy, urlcolor=blue]{hyperref}
\usepackage{subfigure}

\usepackage{graphicx}
\usepackage{txfonts}
\usepackage{lipsum}
\usepackage{subcaption}         
\usepackage{lscape}             
\usepackage{placeins}           
\usepackage{makecell}   
\usepackage{soul}
\sethlcolor{yellow!50}

\begin{document}

   \title{Testing Narrow-jet Gamma-Ray Bursts as Sources of Ultrahigh-Energy Cosmic Rays}


%
%
%

   \author{Saikat Das\inst{1}\corrauth{saikatdas@ufl.edu}        
        \and Soebur Razzaque\inst{2, 3, 4}\email{srazzaque@uj.ac.za}
        \and Justin D. Finke\inst{5,3}\email{justin.d.finke.civ@us.navy.mil}
        \and  Siyao Xu\inst{1}\email{xusiyao@ufl.edu}
        }

   \institute{Department of Physics, University of Florida, Gainesville, FL 32611, USA
   \and Centre for Astro-Particle Physics (CAPP) and Department of Physics, University of Johannesburg, PO Box 524, Auckland Park 2006, South Africa
   \and Department of Physics, The George Washington University, Washington, DC 20052, USA
   \and National Institute for Theoretical and Computational Sciences (NITheCS), Private Bag X1, Matieland, South Africa
   \and U.S.\ Naval Research Laboratory, Code 7653, 4555 Overlook Ave.\ SW,
        Washington, DC,
        20375-5352, USA \\
   }

   \date{Received June 13, 2026}

 
  \abstract
   {Gamma-ray bursts (GRBs) have long been considered candidate sources of ultrahigh-energy cosmic rays (UHECRs) due to their large energy release and relativistic outflows. The detection of multi-TeV $\gamma$-rays from GRB~221009A and its rarity have renewed interest in this connection and motivate considering an additional narrow-jet long GRB population in the local Universe.}
   {We investigate whether such a local narrow-jet population can contribute to the observed diffuse UHECR energy spectrum. We also examine the associated cosmogenic neutrino and cascade $\gamma$-ray emissions to assess the multimessenger viability of this scenario.} 
   {We fit the observed UHECR spectrum and mass composition data using three source-evolution models: a uniform comoving source emissivity, a standard-jet GRB population tracing the star formation rate (SFR), and a standard + narrow-jet GRB population tracing SFR. We propagate a mixed-composition UHECR injection and calculate the cosmogenic neutrino and cascade $\gamma$-ray fluxes.}
   {The standard + narrow jet model fits the observed UHECR spectrum and composition, with the highest-energy flux dominated by the narrow-jet population confined to $z\le z_{\max,*}\simeq0.36$. This low-redshift dominance lowers the cosmogenic neutrino flux compared to the standard-jet GRB population. For the narrow-jet GRB population, the fit implies a baryon loading factor $\xi_{\rm CR}^{\rm nj}\simeq10$.}
   {Such a locally enhanced long-GRB population can therefore explain the highest-energy UHECR flux without violating current multimessenger constraints. Future UHE searches can further probe this scenario through the associated cosmogenic fluxes.}


   \maketitle
\nolinenumbers

\section{Introduction}
Ongoing progress in detecting ultrahigh-energy cosmic rays (UHECRs; $E\gtrsim 10^{18}$ eV) has opened a new window for studying particle acceleration in extreme astrophysical environments \citep[see][for reviews]{Kotera_2011, Anchordoqui:2018qom, AlvesBatista:2019tlv}. The leading experiments observing these highest-energy particles are the Pierre Auger Observatory in Argentina \citep{PierreAuger:2015eyc} and the Telescope Array (TA) in Utah, US \citep{TelescopeArray:2015dcv}. The observed spectrum and mass composition of UHECRs provide important clues to their origin and source environments. The interpretation of the mass composition remains uncertain due to hadronic interaction models extrapolated beyond energies accessible to terrestrial accelerators \citep{2024icrc.confE.365M}, and also differs between Auger and TA. The spectral hardening observed at the ankle, $E\simeq 5\times 10^{18}$ eV, motivates scenarios in which distinct source populations dominate below and above the ankle \citep[e.g.,][]{AlvesBatista:2018zui, Das:2018ymz, Das:2020nvx, Heinze:2019jou}.

A direct association of a UHECR event with an astrophysical source is difficult, since charged particles are deflected by Galactic and intergalactic magnetic fields \citep[see, e.g.,][]{PierreAuger:2017pzq, PierreAuger:2024fgl}. Nearby extragalactic sources may contribute to the observed anisotropy in their arrival directions \citep[e.g.,][]{Liu:2013ppa, Mollerach:2021ifa, Allard:2021ioh, Shaw:2025ykm}. Various candidate source classes have been studied, such as, active galactic nuclei \citep[AGNs; e.g.,][]{Sikora_1987, Dermer_2009NJPh...11f5016D, Dermer_2010ApJ...724.1366D, Essey_2010, Murase_2012, Razzaque_2012, Rodrigues:2020pli, Das:2025tfq}, gamma-ray bursts \citep[GRBs; e.g.,][]{Waxman:1995vg, Vietri:1995hs, Zhang:2003uk, Wang:2007xj, Murase:2008mr, Dermer_2010ApJ...724.1366D, Bustamante:2014oka, Zhang:2017moz, Moore:2023sgo, DeLia:2024kjv}, starburst galaxies \citep{Romero:2018mnb, Attallah:2018euc, Condorelli:2022vfa, Lunardini:2019zcf}, tidal disruption events \citep{Farrar:2014yla, Zhang:2017hom, Plotko:2024gop}, and compact-object mergers \citep{Kotera:2016dmp, Farrar:2024zsm}, although the Galactic to extragalactic transition energy remains uncertain.

Recent multimessenger observations have further sharpened the search for the sources of UHECRs. These include a UHE neutrino candidate of energy 220 PeV detected by KM3NeT \citep{KM3NeT:2025npi, KM3NeT:2025ccp}, and a UHECR event of energy $\approx244$ EeV, reported by the Telescope Array \citep{TelescopeArray:2023sbd}. GRBs have long been considered potential sites of UHECR acceleration because of their large energy release, relativistic outflows, and transient nature. The detection of multi-TeV $\gamma$-rays from the jet of GRB~221009A with an opening angle $\approx0.8^\circ$ \citep{2023ApJ...946L..31B, LHAASO:2023kyg, LHAASO:2023lkv} has led to models in which UHECR-induced cascade emission contributes to the observed very-high-energy ($\epsilon_\gamma\gtrsim0.1$ TeV) $\gamma$-ray signal \citep{AlvesBatista:2022kpg, Das:2022gon, 2023MNRAS.519L..85M}. Motivated by this brightest GRB observed so far and the rarity of such a GRB in the nearby universe \citep{2023ApJ...946L..31B}, a population of narrow-jet long GRBs in the local Universe ($z\lesssim 0.36$) has also been proposed \citep{Finke:2024mni}, in addition to a GRB source evolution model that traces the standard star formation rate (SFR) density \citep{Madau:2014bja}. 

\citet{Finke:2024mni} provide two models. In the ``delta-function'' model, there are two populations, a standard population and a narrow jet population.  All the GRB jets in each population have the same opening angle and energy emitted in $\gamma$-rays ($E_\gamma$), although the populations have different opening angles from each other.  In the more realistic, ``log-normal'' model, for both populations the distributions of opening angles and $E_\gamma$ follow log-normal distributions, with different distributions for the two different populations.  Naturally, in both models, the narrow jet GRB population has smaller opening angles than the standard GRB population. Also in both these models, the narrow-jet distribution must be restricted to low redshift so as not to overproduce the observed rate of faint GRBs.
In this work, we study the implications of a standard + narrow jet GRB source evolution for the UHECR spectrum and composition, together with the associated cosmogenic neutrino ($\nu$) and $\gamma$-ray signals. The standard- and narrow-jet populations differ in their jet opening
angles, local event-rate densities, and redshift evolution. However, we assume the same UHECR injection spectral index, maximum rigidity,
and mass composition for both populations.

Current Auger data suggest a progressively heavier composition above $E\gtrsim10^{18.2}$ eV
\citep{PAO_comp2017}. The distinct source population above the ankle requires a hard injection spectrum $E^{-\alpha}$ with $\alpha\lesssim 1$ \citep{PierreAuger:2016use}. 
Nearby sources ($z\lesssim0.5$) are expected to dominate the highest-energy UHECR flux because nuclei from larger distances are photodisintegrated due to irradiation by photons of energy $\epsilon'_\gamma \sim 8$--$30\,\mathrm{MeV}$ in the nuclear rest frame \citep[see, e.g.,][]{Zhang:2024sjp}. Cosmogenic $\nu$-s and $\gamma$ rays, on the other hand, can receive substantial contributions from higher redshifts. These multimessenger fluxes therefore provide a direct probe of otherwise degenerate UHECR source-evolution models. A local source overdensity is required for the sub-ankle component to satisfy Fermi-LAT diffuse $\gamma$-ray constraints \citep[see, e.g.,][]{Liu:2016brs}. We predict the ``guaranteed'' cosmogenic fluxes under the hypothesis that narrow-jet GRBs are the sources of super-ankle UHECRs.

We describe the source parameters, spectrum and composition fits, and source-evolution models in Sec .~\ref {sec:model}. We present the results in Sec.~\ref{sec:results} and discuss their implications in Sec.~\ref{sec:discussions}.

\section{Model considerations\label{sec:model}}

\subsection{UHECR propagation and simulation setup}
We assume a mixed composition of representative stable nuclei, $^1$H, $^4$He, $^{14}$N, $^{28}$Si, and $^{56}$Fe at injection to fit the spectrum and composition data above the ankle ($E\gtrsim5\cdot10^{18}$ eV), following the approach adopted in the Auger analysis \citep{PierreAuger:2016use}.
All nuclear species are injected following an exponential cutoff power-law spectrum $J(E_i)=\mathcal{N}\sum_A dN_A/dE_i$, where 
\begin{equation}
\dfrac{dN_A}{dE_i}
=
\left(\dfrac{E_i}{E_0}\right)^{-\alpha}
\begin{cases}
1, & E_i\leq Z_A R_{\rm cut},\\
\exp\left(1-{E_i}/{Z_A R_{\rm cut}}\right), & E_i>Z_A R_{\rm cut},
\end{cases}
\label{eq:injection}
\end{equation}
where $f_A$ is the injected abundance fraction of species with mass $A$, $Z_A$ is its charge, $\alpha$ is the injection spectral index, and $R_{\rm cut}$ is the maximum rigidity. The constants $\mathcal{N}$ and $E_0$ are an arbitrary normalization and reference energy, respectively. A single-population fit across the ankle is viable primarily in the
proton-dip model \citep{Berezinsky:2002nc}, but this interpretation is strongly disfavored by Auger composition measurements \citep{PierreAuger:2016qzj} and neutrino fluxes \citep{IceCube:2016uab}. Hence the ankle here is interpreted as a transition between two (or more) distinct source populations.

UHECRs interact with the cosmic microwave background (CMB) and the extragalactic background light (EBL) during their propagation over cosmological distances. We use the \texttt{CRPropa 3.2} Monte Carlo simulation framework to find the particle yields obtained at Earth after extragalactic propagation \citep{AlvesBatista:2016vpy, AlvesBatista:2022vem}. We include all energy-loss processes for UHECRs, namely photopion production, Bethe-Heitler pair production, photodisintegration, nuclear beta decay, and adiabatic expansion of the Universe, in addition to energy loss of secondary $e^\pm$ and $\gamma$-rays. We find the best-fit values of the UHECR parameters $\alpha$, $R_{\rm cut}$, and $f_A$ for the source evolution models studied in this work. We consider UHECR injection over a distance of $1-5000$ Mpc and in the energy range $0.1-1000$ EeV. 
The spectrum of EBL photons and its evolution with redshift are not as well known as those of the CMB. We use the \citet{Gilmore_2012} EBL model and \texttt{TALYS} 1.8 photodisintegration cross-section \citep{Koning_2005}.

\subsection{Source Redshift Evolution\label{sec:source}}

The comoving rate evolution of standard-jet GRBs is assumed to trace the comoving star-formation rate (SFR) density $\psi(z)$ \citep{Madau:2014bja}, where $z$ is the source redshift. The evolution function can be expressed in a dimensionless form, normalized to unity at $z=0$ as
\begin{equation}
\mathcal{H}_{\rm sj}(z) = \frac{\psi(z)}{\psi(0)},
\qquad
\text{where,}\, \, \, 
\psi(z) = \frac{a_s(1+z)^{b_s}}{1+\left[(1+z)/c_s\right]^{d_s}}.
\end{equation}
The best-fit parameters obtained in \citet{Finke_2022} are $a_s=9.2\times10^{-3}\ M_\odot\ {\rm yr^{-1}\ Mpc^{-3}}$, $b_s=2.79$, $c_s=3.10$, and $d_s=6.97$. 
We adopt the narrow-jet GRB population proposed by \citet{Finke:2024mni}.
In the simple delta-function model, each GRB is described by a single jet opening angle and $\gamma$-ray energy.
Applied to GRB 221009A, this motivates a nearby narrow-jet population with jet opening angle $\hat\theta_j=0.8^\circ$ and $\hat E_\gamma/\hat\theta_j^2\simeq 1.7\times10^{51}\ {\rm erg\ deg^{-2}}$, for a beaming-corrected $\gamma$-ray energy $E_\gamma^{\rm nj} \equiv {\hat E_\gamma} \simeq10^{51}\ {\rm erg}$. To avoid overproducing the observed rate predicted from standard jet GRBs, the narrow-jet population is restricted to the nearby Universe, with source evolution 
$\mathcal{H}_{\rm nj}(z)
\propto
\psi(z)\Theta(z_{\max,*}-z)$,
where $\Theta$ is the Heaviside step function, imposing a cutoff at $z_{\max,*}\simeq0.36$. In this model, the rate normalizations of the standard and narrow-jet GRB populations are denoted by $K_n$ and $K'_n$, respectively. In the more realistic log-normal population model, which allows distributions in
$E_\gamma$ and $\theta_j$, the fitted values from \citet{Finke:2024mni} are $K_n=10^{-6}\ M_\odot^{-1}$ and $K'_n=4\times10^{-6}\ M_\odot^{-1}$. 
The effective standard + narrow jets evolution in dimensionless form and normalized to unity at $z=0$ can be written as
\begin{equation}
    \mathcal{H}_{\rm tot}(z) = \mathcal{H}_{\rm sj}(z) + \mathcal{H}_{\rm nj}(z) =\frac{\psi(z)}{\psi(0)}\left[\frac{1+(K^\prime_n/K_n)\Theta(z_{\max,*}-z)}{1+K^\prime_n/K_n}\right]
\end{equation}
%
The narrow-jet GRB event rate per unit comoving volume is $\dot{n}_{\rm nj}(z)=\dot{n}_{\rm nj}(0)\mathcal{H}_{\rm nj}(z)$,
where $\dot n_{\rm nj}(0)\simeq4\times10^{-8}\ {\rm Mpc^{-3}\ yr^{-1}}$. 
This corresponds to a local narrow-jet $\gamma$-ray energy-injection rate density of $\dot n_{\rm nj}(0)E_\gamma^{\rm nj} \simeq 4\times10^{43}\ {\rm erg\ Mpc^{-3}\ yr^{-1}}$.

\subsection{Composition and shower depth distribution}

The mean shower-depth maximum, $\langle X_{\rm max}\rangle$, and its dispersion are computed from the first two moments of $\ln A$ using the Auger parametrization based on the Heitler model of extensive air showers \citep{Matthews_2005, PierreAuger:2013xim},
\begin{align}
\langle X_{\rm max} \rangle = \langle X_{\rm max} \rangle_p + f_E \langle \ln A \rangle \\
\sigma^2 (X_{\rm max}) = \langle \sigma^2_{\rm sh} \rangle + f_E^2 \sigma^2_{\ln A}
\end{align}
where $\langle X_{\rm max} \rangle_p$ is the mean maximum shower depth of protons and the parameter $f_E$ depends on the UHECR energy,
\begin{equation}
f_E = \xi - \dfrac{D}{\ln 10} + \delta\log_{10}\bigg(\dfrac{E}{E_0}\bigg)
\end{equation} 
where $\xi$, $D$, and $\delta$ depend on the specific hadronic interaction model. $\sigma^2_{\ln A}$ is the variance of $\ln A$ distribution and $\langle \sigma_{\rm sh}^2 \rangle$ is the average variance of $X_{\rm max}$ weighted by the $\ln A$ distribution,
\begin{equation}
\langle \sigma_{\rm sh}^2 \rangle = \sigma_p^2[1+ a \langle \ln A \rangle + b \langle (\ln A)^2 \rangle]
\end{equation}
where $\sigma_p^2$ is the $X_{\rm max}$ variance for proton showers depending on energy and three model-dependent parameters. In this work, we use the updated parameter values\footnote{S. Petrera and F. Salamida (2018), Pierre Auger Observatory} obtained from the \texttt{Conex} simulations \citep{Pierog:2004re}, for one of the post-LHC hadronic interaction models, \texttt{Sybill2.3c} \citep{Riehn:2015oba}.

\section{Results\label{sec:results}}

We fit the observed UHECR spectrum and composition for three source-evolution models: (i) a uniform source distribution corresponding to constant comoving emissivity, $\mathcal{H}(z)=1$, (ii) a standard-jet GRB population tracing the SFR evolution, and (iii) standard + narrow jet GRB population, also tracing the SFR evolution (see Table~\ref{tab:best_fit_param}). In each case, the injected abundance fractions, spectral index, rigidity cutoff, and normalization are varied to reproduce the Auger spectrum and the first two moments of the inferred mass composition. 

\begin{figure}
    \centering
    \includegraphics[width=0.46\textwidth]{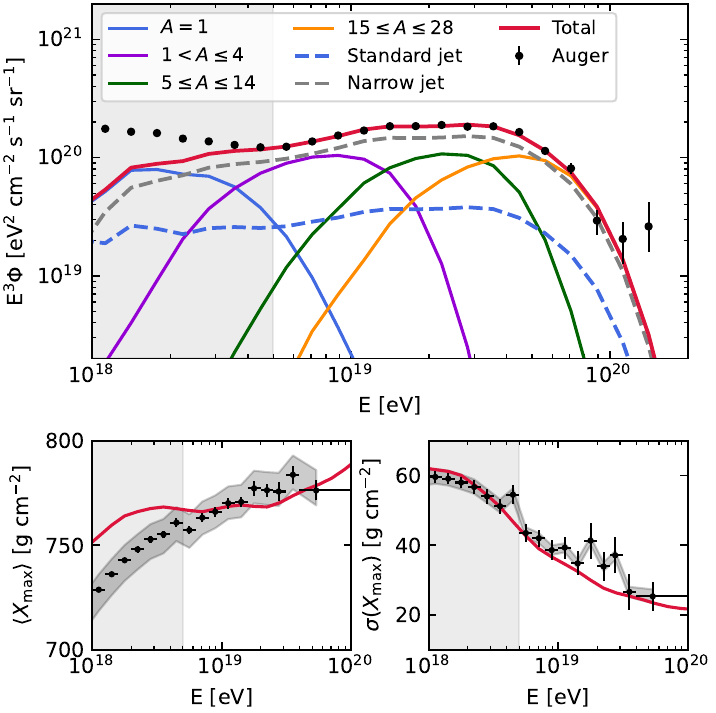}
    \caption{\small{Combined fit to UHECR spectrum and composition data from Auger \citep{PierreAuger:2018gfc, PierreAuger:2020kuy, PierreAuger:2020qqz} using a standard + narrow jet GRB population model. The energy range excluded from the fit is shown by grey shading. Note that the abundance fraction of Fe is zero at injection.}}
    \label{fig:sfr_plus_narrow}
\end{figure}

The best-fit result for the standard + narrow jet GRB evolution described in Sec.~\ref{sec:source} is shown in Fig.~\ref{fig:sfr_plus_narrow}, together with the corresponding $\langle X_{\rm max}\rangle$ and $\sigma(X_{\rm max})$ fits. 
Their individual contributions are shown by the blue and gray dashed curves. We assume the same injected mass composition for both populations, and show the resulting abundances at Earth for the total flux. At the highest energies, the flux is dominated by the nearby narrow-jet population, which is restricted to $z_{\max,*}\simeq0.36$ and is therefore less affected by photodisintegration during propagation.
\begin{table}[ht!]
\caption{\label{tab:best_fit_param}UHECR spectrum and composition best-fit parameters}
\centering
\begin{tabular}{cccc}
\hline\hline
& \makecell{No\\ evolution} & \makecell{Standard\\ jet only} & \makecell{Standard + \\Narrow jets}\\
\hline
Spectrum, $\alpha$ & $-1.3$ & -1.8 & -1.6 \\
$\log_{10}(R_{\rm cut}/V)$ & 18.2 & 18.2 & 18.2 \\
\hline
$f_H$ (\%)    & 68.0   & 39.0  & 65.0 \\
$f_{He}$ (\%) & 31.0   & 59.0  & 34.0 \\
$f_N$ (\%)    & 1.0    & 2.0   & 1.0  \\
$f_{Si}$ (\%) & 0.03   & 0.06  & 0.04 \\
$f_{Fe}$ (\%) & 0.0    & 0.0   & 0.0  \\
$\delta_E$    & -0.09  & 0.08  & 0.06 \\
$\chi^2/{\rm d.o.f}$ & 1.73 & 1.98 & 1.94 \\
\hline
\makecell{$\dot \varepsilon_{\rm CR}(z=0)$\\ erg Mpc$^{-3}$ yr$^{-1}$} & $4.7\times10^{44}$ & $5.2\times10^{44}$ & $5.1\times10^{44}$ \\
\hline
\end{tabular}
\end{table}

The goodness-of-fit is calculated using the $\chi^2$ statistic,
\begin{align}
\chi^2_j = \sum_k \dfrac{[J_k^{\rm sim}(E'_k; f_k)-J_k^{\rm ob}(E_k)]^2}{\sigma_k^2} + \bigg(\dfrac{\delta_E}{\sigma_E}\bigg)^2
\end{align}
where the summation runs over all energy bins $k$ included in the fitting procedure and $j$ corresponds to each of the three observables, viz., the energy spectrum, $X_{\rm max}$, and $\sigma(X_{\rm max})$. The systematic error in the Auger spectrum data is dominated by the 14\% energy uncertainty $\sigma_E$. We introduce a nuisance parameter $\delta_E$ such that $E'_i = (1+\delta_E)E_i$, where $\delta_E$ is varied in the range $-0.14\leqslant\delta_E\leqslant+0.14$ to find the lowest $\chi^2$. The flux normalization of the simulated spectrum is also treated as a free parameter. 

The best-fit source parameters are listed in Table~\ref{tab:best_fit_param}, together with the minimum $\chi^2$ per degree of freedom (d.o.f). 
For all source-evolution models, the injected composition is dominated by light nuclei
while the propagated flux becomes
heavier above the ankle because of the rigidity-dependent cutoff in the injection spectrum. We scan $\alpha\in[-2.0,0]$ and $\log_{10}(R_{\rm cut}/{\rm V})\in[17.8,19.0]$ with grid spacing $0.1$ in both parameters. The H, He, and N fractions are varied in integer percentage
steps, while Si and Fe are varied with a stepsize of $0.1\%$ because of their small contributions.

The adopted source evolution model affects the fitted normalization. We write the redshift dependence of the differential comoving energy injection rate density of nuclei with mass A as
\begin{align}
    \mathcal{Q}_A(E_i,A,z)
=
\mathcal{N}\,
\frac{dN_A}{dE_i}\,
\mathcal{H}(z),
\end{align}
where the nuclear abundance $f_A$ is absorbed in the injection spectrum ${dN_A}/{dE_i}$ (see Eqn.~\ref{eq:injection}).
The corresponding UHECR energy-injection rate density ($E\gtrsim10^{17}$ eV) at redshift $z$ is
\begin{equation}
\dot{\varepsilon}_{\rm CR}(z)
=
\sum_A
\int_{E_{i,\, \rm min}}^{E_{i,\,\rm max}}
dE_i\,
E_i\,
\mathcal{Q}_A(E_i, A,z),
\end{equation}
where the integration is over the energy range $0.1-1000$ EeV. The local value, $\dot{\varepsilon}_{\rm CR}(0)$, is obtained by evaluating this expression at $z=0$ and note that $\mathcal{H}(z=0)$ is normalized to unity. The values of $\dot{\varepsilon}(0)$ for the different models are presented in Table~\ref{tab:best_fit_param}.

Since $K'_n/K_n=4$ for the standard + narrow jet evolution model, the narrow-jet component contributes a fraction $f_{\rm nj}=0.8$ of the local UHECR energy-rate density. The corresponding narrow-jet GRB cosmic-ray loading factor is
    $\xi_{\rm CR}^{\rm nj} = {\dot{\varepsilon}_{\rm CR}^{\rm nj}(0)} / {\dot n_{\rm nj}(0)E_{\gamma}^{\rm nj}}= {f_{\rm nj}\dot{\varepsilon}_{\rm CR}^{\rm total}(0)}/{\dot n_{\rm nj}(0)E_{\gamma}^{\rm nj}}$
where $\dot{\varepsilon}_{\rm CR}^{\rm total}(0)$ is the total local UHECR energy injection rate obtained from the standard + narrow jet fit, $\dot n_{\rm nj}(0)$ is the intrinsic local rate density of narrow-jet GRBs, and $E_{\gamma}^{\rm nj}$ is the beaming-corrected gamma-ray energy per burst. Thus, the UHECR fit constrains the product
$\dot n_{\rm nj}(0)\xi_{\rm CR}^{\rm nj}$. Using the values $K'_n/K_n=4$, a rate of $\dot n_{\rm nj}(0)=4\times10^{-8}$ Mpc$^{-3}$ yr$^{-1}$, and $E_{\gamma}^{\rm nj}=10^{51}\ {\rm erg}$, the required loading factor is
\begin{align}
\xi_{\rm CR}^{\rm nj} \simeq 10\, \times &
\left(\frac{\dot{\varepsilon}_{\rm CR}^{\rm total}(0)}{5\times 10^{44}\ {\rm erg\ Mpc^{-3}\ yr^{-1}}}\right)
\left(\frac{f_{\rm nj}}{0.8}\right) \nonumber \\
&\times\left(\frac{\dot n_{\rm nj}(0)}{4\times10^{-8}\ {\rm Mpc^{-3}\ yr^{-1}}}\right)^{-1}
\left(\frac{E_{\gamma}^{\rm nj}}{10^{51}\ {\rm erg}}\right)^{-1},
\end{align}
consistent with the range estimated in \citet{Finke:2024mni}. Conversely, the required local narrow-jet GRB rate can be inferred for an assumed $\xi_{\rm CR}^{\rm nj}$. For comparison, the standard-jet-only fit to UHECR data requires a baryon-loading factor of $\sim50$ derived from a similar calculation.

The cosmogenic $\nu$ and $\gamma$-ray spectra are calculated using the same best-fit UHECR parameters obtained from the spectrum and composition fits. We perform 1D simulations for each injected nuclear species to track secondary production. The $\nu$ yields at Earth are obtained directly from these simulations.
Secondary $e^\pm$ and $\gamma$-rays produced by UHECR interactions with the CMB and EBL, as well as $\beta$-decay of secondary neutrons, initiate electromagnetic cascades down to GeV-TeV energies through pair production, inverse-Compton emission, and synchrotron radiation in cosmic magnetic fields.
The electromagnetic secondaries produced during UHECR propagation are stored at production and subsequently propagated by solving the 1D transport equations following \citet{Lee:1996fp}. The resulting $\nu$ and cascade $\gamma$-ray spectra at Earth are obtained by integrating the output flux from redshift bins after convolution with the source-evolution function $\mathcal{H}(z)$, with each injected species weighted by its best-fit abundance and rigidity-dependent injection spectrum. The fluxes are normalized to the corresponding local UHECR energy-generation-rate density as presented in Table~\ref{tab:best_fit_param}. Thus the cosmogenic fluxes at Earth are obtained as
\begin{align}
\Phi_{\nu,\gamma}(E)
=
\frac{c}{4\pi}\,
\mathcal{N}
\sum_A f_A &
\int_{z_{\min}}^{z_{\max}} dz\,
\left|\frac{dt}{dz}\right|
\mathcal{H}(z) \nonumber \\
&\times\int_{E_{i,\, \rm min}}^{E_{i,\,\rm max}}
dE_i\,
\frac{dN_A}{dE_i}\,
Y_A^{\nu,\gamma}(E,E_i,z),
\end{align}
where $Y_A^{\nu, \gamma}$ is the propagated spectrum at Earth from monoenergetic UHECR injection at redshift $z$ and energy $E_i$. The cosmological time-redshift Jacobian is
\begin{equation}
\left|\frac{dt}{dz}\right|
=
\frac{1}{H_0(1+z)
\sqrt{\Omega_m(1+z)^3+\Omega_\Lambda}},
\end{equation}
where a flat $\Lambda$CDM cosmology is assumed, with $H_0 = 67.3$~km s$^{-1}$ Mpc$^{-1}$ and $\Omega_m = 0.315$. 

\begin{figure}
    \centering
    \includegraphics[width=0.46\textwidth]{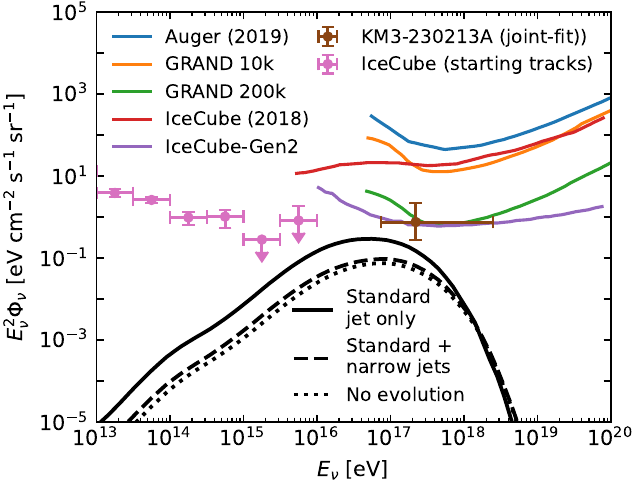}
    \caption{\small{All-flavor cosmogenic $\nu$ fluxes for the models presented in Table~\ref{tab:best_fit_param}. The flux upper limits from the IceCube \citep{IceCube:2018fhm}, Auger \citep{PierreAuger:2019ens}, and the projected 90\% C.L. sensitivities of the IceCube-Gen2 radio array \citep{IceCube:2019pna, IceCube-Gen2:2020qha}, GRAND-10k, and GRAND200k \citep{GRAND:2018iaj} are also shown.}}
    \label{fig:cosmogenic_neutrino}
\end{figure}

The all-flavor cosmogenic $\nu$ spectra are shown in Fig.~\ref{fig:cosmogenic_neutrino} for the models in Table~\ref{tab:best_fit_param}. The predicted fluxes are consistent with the IceCube diffuse spectrum and remain below current differential flux upper limits. The standard jet-only model gives the highest flux, peaking at $E_\nu\approx10^{17}$ eV. Although the fitted $f_H$ is lower for the latter model, its stronger high-redshift evolution yields a higher cosmogenic $\nu$ flux. For the standard + narrow jets, the reduction follows from the low-redshift cutoff of the narrow-jet component, $z\le z_{\max,*}\simeq0.36$, which reduces the propagation distance and hence the $p\gamma$ interaction probability. Moreover, protons are the most efficient producers of cosmogenic $\nu$-s, whereas heavier nuclei lose most of their energy through photodisintegration. The constraints are particularly sensitive to strongly evolving, proton-rich UHECR source populations with high maximum energies \citep[see, e.g.,][]{Sherman:2025gjn}. 

The flux level of the UHE neutrino event KM3-230213A is also shown, using the joint-fit estimate based on IceCube, Auger, and KM3NeT data. If attributed to a diffuse cosmogenic neutrino flux \citep[see, e.g.,][]{KM3NeT:2025vut, Kuznetsov:2025ehl, Das:2025vqd, Boxi:2025ony, Alhebsi:2026bdk}, the inferred flux level is compatible with the source-evolution models considered here.
The peak neutrino flux from the standard+narrow-jet model lies within an order of magnitude of the KM3-230213A estimate. Previous work showed that this level can be reproduced by diffuse astrophysical neutrinos from GRB afterglow \citep{Razzaque2013PhRvD..88j3003R, Razzaque2015PhRvD..91d3003R}. For an ISM density $n_0\sim1~{\rm cm}^{-3}$, this corresponds to a baryon loading $\xi_{\rm CR}\leq 27$ at 68\% C.L. \citep{KM3NeT:2025zmb}. In our cosmogenic scenario, however, the mixed composition leads to lower rigidities and suppresses the GZK neutrino flux. The IceCube starting-track data provide a measurement of the diffuse astrophysical neutrino flux in the TeV--PeV range, serving as a low-energy consistency check \citep{IceCube:2024fxo}. 

Protons are the most efficient producers of cosmogenic $\gamma$ rays, through $\pi^0$-decay photons. In contrast, the contribution from heavier nuclei is dominated by photodisintegration and inverse-Compton emission from secondary electrons. 
Pure-proton models with strong source redshift evolution can therefore exceed the Fermi-LAT measurement of the isotropic $\gamma$-ray background \citep[IGRB;][]{Fermi-LAT:2014ryh}, and are strongly constrained \citep{Berezinsky_2011}. 
For weak intergalactic fields $B\lesssim 10^{-12}\,{\rm G}$, synchrotron losses remain negligible.

\begin{figure}
    \centering
    \includegraphics[width=0.46\textwidth]{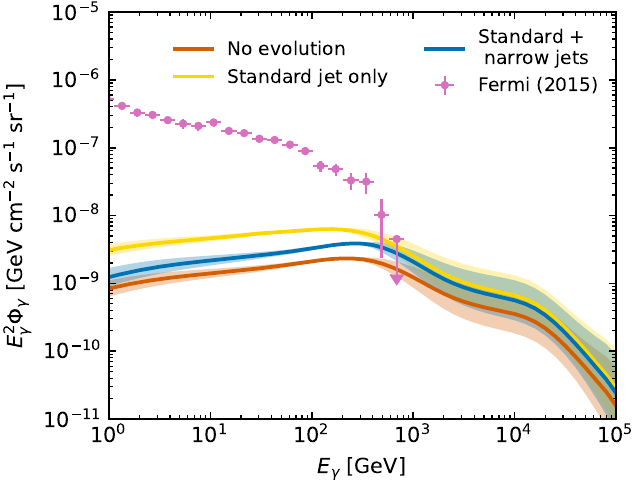}
    \caption{\small{Cosmogenic $\gamma$-ray fluxes for different source redshift evolution models, assuming $z_{\rm max}=2$. 
    Shaded bands indicate the uncertainty from the infrared and radio backgrounds used in the cascade calculation. The Fermi-LAT IGRB flux is shown for Galactic foreground model A \citep{Fermi-LAT:2014ryh}.}
    }
    \label{fig:gamma_flux}
\end{figure}

Figure~\ref{fig:gamma_flux} shows the cosmogenic $\gamma$-ray fluxes corresponding to the best-fit UHECR models. 
The shaded bands indicate the uncertainty from different infrared-background normalizations \citep{Franceschini:2008tp,Finke_2010} and from including or excluding the radio background of \citet{Protheroe:1996si}.
The predicted $\gamma$-ray fluxes for the best fit parameters presented in Table~\ref{tab:best_fit_param} saturate the Fermi-LAT upper limit at $\sim1$ TeV. 
Although the fitted $f_H$ is lower for the standard-jet-only model, its stronger evolution increases the production of high-energy photons at large redshifts, which are reprocessed, thereby enhancing the sub-TeV flux. In addition, the rigidity-dependent cutoff in the injection yields a non-negligible contribution from heavier nuclei. While the standard-jet-only population saturates the upper limit near $\sim1$ TeV, the standard + narrow jet population allows contribution from other sources such as $\gamma$-ray blazars. 
For standard + narrow jets, the highest-energy UHECR contribution comes from the low-redshift narrow-jet component, which also dominates the $\gamma$-ray flux. 




\section{Discussions and Conclusions\label{sec:discussions}}

The upcoming AugerPrime \citep{Castellina:2019irv} and TA$\times4$ \citep{TelescopeArray:2021dri} upgrades will improve measurements of the UHECR spectrum, composition, and anisotropy, including possible intermediate-scale anisotropies or correlations with nearby transient populations \citep{TelescopeArray:2018rtg, PierreAuger:2018qvk, Marafico:2024qgh}. Multimessenger searches for neutrino counterparts to long GRBs will provide a complementary test of the GRB source scenario.
The source redshift evolution is difficult to determine from UHECR spectrum and composition data alone due to degeneracies among model parameters.

Cosmogenic $\gamma$-ray and neutrino constraints already disfavor strong source evolution, $\gtrsim (1+z)^3$, and proton-dip scenarios \citep{Aloisio:2006wv}. However, in a mixed-composition scenario, the current Auger fits favor very hard injection spectra. Depending on the assumed propagation model, EBL model, and photodisintegration cross sections, the best-fit spectral indices can even become negative \citep{PierreAuger:2016use}, which is consistent with the values obtained here. However, it is difficult to reconcile such a hard injection spectrum with standard acceleration models \citep[see, e.g.,][]{Globus_2025}. Thus, the fitted $\alpha$ values obtained here can be interpreted as the spectral indices of particles escaping the source environment. UHECR acceleration by internal shock in GRBs predicts an escaping spectral index $\alpha\approx0$ \citep{Baerwald:2013pu}. Such a hard spectral index is often interpreted as the result of enhanced interactions in the source environment \citep{Unger:2015laa, Kachelriess:2017tvs, Supanitsky:2018jje, Muzio:2019leu}.


The cosmogenic $\nu$ flux in our models remains below the reach of current and planned detectors. KM3NeT in the northern hemisphere and the next-generation IceCube radio array, with an effective volume about ten times larger than IceCube, will provide stronger constraints on such source models.  Our transport-equation calculation for $\gamma$-rays differs from Monte Carlo particle tracking, and uncertainties in the flux arise from discrepancies in energy conservation and various numerical methods employed \citep[see, e.g.,][]{Murase:2015xka, Kalashev:2022cja}. For both $\nu$ and $\gamma$-rays, an additional flux suppression arises from the low rigidity cutoff, below GZK energies and heavy nuclei dominance at the highest energies. This also implies that the cutoff in the highest-energy UHECR spectrum is due to the maximum acceleration energy at the sources, rather than the GZK effect \citep{Greisen1966PhRvL..16..748G, Zatsepin1966JETPL...4...78Z}.

Current neutrino upper limits during prompt emission remain compatible with the baryon loading obtained here, for suitable GRB parameters and emission scenarios \citep{Zhang_2011, Zhang:2012qy, IceCube:2016ipa,IceCube:2017amx}. It has been shown that UHE nuclei accelerated in GRB internal shocks can survive photodisintegration for sufficiently large dissipation radii or bulk Lorentz factors. Survival is also possible if the prompt-emission self-absorption break lies above several keV \citep{Wang:2007xj}. Moreover, prompt-neutrino constraints are weaker for sufficiently large dissipation radii or Lorentz factors \citep{Murase:2022vqf}, while multizone models allow neutrino emission and UHECR escape to occur at different collision radii, indicating significantly milder dependence on baryon loading \citep{Bustamante:2014oka}. Some low-luminosity GRBs may host successful jets with relatively low Lorentz factors \citep[e.g.,][]{Zhang_2012}, in which internal shocks can operate and accelerate UHECRs \citep{Zhang:2017moz}. However, if the electromagnetic emission is produced from trans-relativistic shock breakout, internal shocks may be absent \citep{Nakar_2012}. Survival of UHECR nuclei in an external shock during the afterglow is naturally possible due to a large dissipation radius and a lower radiation density. We infer a baryon-loading factor for the narrow-jet population, consistent with KM3NeT expectations \citep{KM3NeT:2025zmb, KM3NeT:2024nwb}. Afterglow modeling indicates that only a small fraction of the shock energy goes into electrons ($\sim0.01-0.1$), implying that substantially more energy can remain available for cosmic-ray acceleration than is carried by the radiating electrons \citep{Barnard:2025zjx}.

A local narrow-jet long-GRB population can explain the UHECR spectrum and composition data while remaining consistent with current cosmogenic $\gamma$-ray and neutrino upper limits. 
The observation of multi-TeV $\gamma$-rays from extragalactic transients may indicate a cosmogenic origin. Such scenarios have been invoked for blazars \citep{Essey_2010}. Next-generation imaging atmospheric Cherenkov telescopes such as LHAASO and CTA \citep{Gueta:2021vO} may be able to resolve such a component in GRBs \citep[see][]{Das:2022gon}. Future ground-based $\gamma$-ray detectors such as SWGO \citep{Conceicao:2023tfb} may increase the discovery space for very-high-energy emission from exceptionally bright GRBs \citep[see, e.g.,][]{Huang_2026EPJC...86..195H}, although establishing a rare narrow-jet population may require very long timescales \citep{Finke:2024mni}. In the longer term, nondetections of the diffuse cosmogenic neutrino flux by next-generation UHE neutrino detectors such as the IceCube-Gen2 radio array, GRAND, and PUEO \citep{PUEO:2020bnn} would constrain the contribution of GRBs to the UHECR spectrum.

\begin{acknowledgements}
    S.D. and S.X. acknowledge the support from the NASA ATP award 80NSSC24K0896 and the NASA Heliophysics Living with A Star Science Program 80NSSC25K0067.
    S.R. was partially supported by a BRICS STI grant and by a NITheCS grant from the National Research Foundation, South Africa. J.D.F.\ is supported by the Office of Naval Research. The authors acknowledge UFIT Research Computing for providing computation resources and support that have contributed to the research results reported in this article.
\end{acknowledgements}

\bibliographystyle{aa} 
\bibliography{aa.bib}

\begin{appendix}

\end{appendix}

\end{document}